\documentclass[prl,showpacs,amsmath,twocolumn]{revtex4}  %% REVTeX 4.0
\usepackage{graphicx}% Include figure files
\usepackage{amssymb}
\usepackage{bm}

\begin{document}
\title{
Continuous macroscopic limit of a discrete stochastic
 model\\ for interaction of living cells}

\author{Mark Alber$^{1}$ }
\email{malber@nd.edu}
\author{Nan Chen$^{1}$}
\author{Pavel M. Lushnikov$^{2,3}$}
\author{Stuart A. Newman$^{4}$} \affiliation{$^1$Department of
Mathematics, University of Notre Dame, Notre Dame,
46656 \\
$^2$ Department of Mathematics and Statistics, University of New
Mexico,
Albuquerque, NM 87131, USA \\
  $^3$ Landau Institute for Theoretical Physics, Kosygin St. 2,
  Moscow, 119334, Russia\\
  $^4$ Department of Cell Biology and Anatomy, New York Medical
  College, Valhalla, NY 10595, USA}

\date{\today }

\begin{abstract}
In the  development of multiscale biological models it is crucial
to establish a connection between discrete microscopic or
mesoscopic stochastic models and macroscopic continuous
descriptions based on cellular density. In this paper a continuous
limit of a two-dimensional Cellular Potts Model (CPM) with
excluded volume is derived, describing cells moving in a medium
and reacting to each other through both direct contact and long
range chemotaxis. The continuous macroscopic model is obtained as
a Fokker-Planck equation describing evolution of the cell
probability density function.  All coefficients of the general
macroscopic model are derived from parameters of the CPM and a
very good agreement is demonstrated between CPM Monte Carlo
simulations and numerical solution of the macroscopic model. It is
also shown that in the absence of contact cell-cell interactions,
the obtained model reduces to the classical macroscopic
Keller-Segel model. General multiscale approach is demonstrated by
simulating spongy bone formation from loosely packed mesenchyme
via the intramembranous route suggesting that self-organizing
physical mechanisms can account for this developmental process.
\end{abstract}

\pacs{ 87.18.Ed, 05.40.Ca, 05.65.+b, 87.18.Hf, 87.18.Bb; 87.18.La;
87.10.1e}

\maketitle

%\thanks{This work was partially supported by NIH Grant No. 1R0-GM076692-01:
%Interagency Opportunities in Multiscale Modeling in Biomedical,
%Biological and Behavioral Systems NSF 04.6071. Simulations were
%performed on the Notre Dame Biocomplexity Cluster supported in part
%by NSF MRI Grant No. DBI-0420980.}


\date{Dec. 25, 2006}

%%% ----------------------------------------------------------------------

%%% ----------------------------------------------------------------------
%\maketitle
%%% ----------------------------------------------------------------------
%\tableofcontents

A large literature exists studying continuous limits of point-wise discrete
microscopic models for biological systems. For example, the classic Keller-Segel
PDE model of chemotaxis [1] was derived
from a discrete model with point-wise cells undergoing random walk [2-5].
However, many biological phenomena require taking into account the finite
size of biological cells, and much less work has
been done on deriving macroscopic limits of microscopic models which treat
cells as extended objects. The mesoscopic Cellular Potts Model (CPM), first
introduced by Glazier and Graner [6, 7], has
been used as a component of multiscale, experimentally motivated hybrid approaches,
combining discrete and macroscopic continuous representations,  to simulate,
among others, morphological phenomena
in the cellular slime mold Dictyostelium discoideum \cite{Pa}, vascular
development \cite{Merks} and the proximo-distal increase in the number of
skeletal elements in the developing avian limb
\cite{Fram1}.

One of the earliest attempts at combining mesoscopic and macroscopic
levels of description of cellular dynamics was described in
\cite{turner2004} where the diffusion coefficient for a collection
of noninteracting randomly moving cells was derived from a
one-dimensional CPM. Recently a microscopic limit of subcellular
elements model \cite{NewmanMathBioEng2005} was derived in the form
of an advection-diffusion partial differential equation for cellular
density. In previous papers \cite{alber1, alber2} we studied the
continuous limit of 1D and 2D models of individual cell motion in a
medium, in the presence of an external field but without contact
cell-cell  interactions.

\begin{figure}
\begin{center}
\includegraphics [width=40mm]{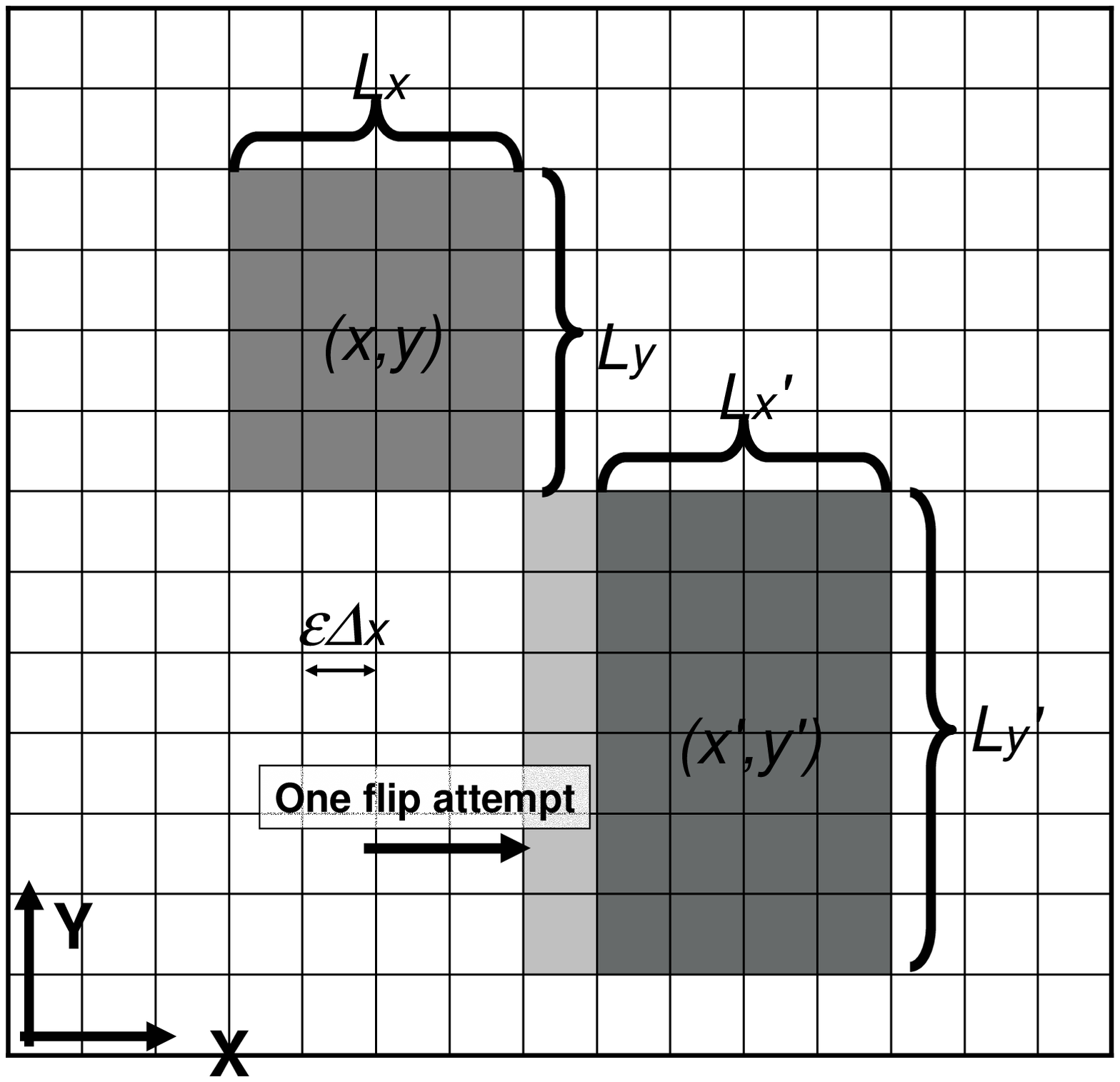}
\caption{Cell representation in the two dimensional CPM. In this
picture grey and white colors are used to indicate the cell body
and ECM respectively. Cell can grow or shrink in $x$ and $y$
direction by adding or removing one row (or column) of
pixels.}\label{figMonte}
\end{center}
\end{figure}
This paper describes a theoretical analysis leading to a
continuous macroscopic limit of the two-dimensional mesocopic CPM
with contact cell-cell  interactions. Our approach, which is based
on combining mesoscopic and macroscopic
 models, can be applied to studying biological phenomena in which
 a nonconfluent
population of cells interact directly and via soluble factors,
forming an open network structure. Examples include vasculogenesis
\cite{Merks,rupp,szabo,gamba} and formation of trabecular, or
spongy, bone \cite{cormack,courtin,tabor} to be described below.

The CPM, defined on a multidimensional lattice, allows simulation
of both cell-cell contact and chemotactic long distance
interactions, along with extended cell representations. In
deriving below our continuous model we assume that cells interact
with one another subject to an excluded volume constraint. In the
CPM a multidimensional integer index is associated with each
lattice site (\emph{pixel}) to indicate that a pixel belongs to a
cell of particular type or medium. Each cell is represented by a
cluster of pixels with the same index. Pixels evolve according to
the classical Metropolis algorithm based on Boltzmann statistics,
and the effective energy
\begin{equation}
E = E _{Adhesion} + E _{Perimeter} + E _{Field}.\label{eq:2}
\end{equation}
Namely, if a proposed change in a lattice configuration results in
energy change $\Delta E$, it is accepted with probability
\begin{equation}\label{Phi}
\Phi(\Delta E) = \left \{\begin{array}{cc}1, & \Delta E \leq 0\\
e^-\frac{\Delta E}{T}, &  \Delta E>0, \end{array} \right.
\end{equation}
where $T$ represents an effective boundary fluctuation amplitude
of model cells in units of energy. Since the cells' environment is
highly viscous, cells move to minimize their total energy
consistent with imposed constraints and boundary conditions. If a
change of a randomly chosen pixels' index causes cell-cell overlap
it is abandoned. Otherwise, the acceptance probability is
calculated using the corresponding energy change. If the change
attempt is accepted, this results in changing location of the
center of mass and dimensions of the cell.

In this paper we assume that each cell has a rectangular shape,
that it moves or changes its shape by adding or removing a row or
column of pixels (see Figure \ref{figMonte}) and that cells come
into direct contact with each other. They also interact with each
other over long distances through producing diffusing chemicals
and reacting to local chemical gradients  (process called
chemotaxis). Although we model adhesion between cells and the
extracellular matrix (ECM), we neglect cell-cell adhesion and take
into account cell-cell interaction from excluded volume constraint
meaning that cells cannot occupy the same volume. Under these
assumptions terms in the Hamiltonian (\ref{eq:2}) have the
following forms. $E _{Adhesion}$ phenomenologically describes the
net adhesion or repulsion between the cell surface and ECM and it
is a product of the binding energy per unit length, $J_{CM}$, and
the length of an interface between the cell boundary and ECM: $
E_{Adhesion}=2J_{cm}(L_x+L_y)$.  $E_\mathit{Perimeter}$ defines an
energy penalty function for dimensions of a cell deviating from
the target values  $L_{T_x}$ and $L_{T_y}$: $ E _{Perimeter} =
\lambda_x(L_x-L_{T_x})^2+\lambda_y(L_y-L_{T_y})^2$ where
$\lambda_x$ and $\lambda_y$ are constants.  Cells can move up or
down gradients of both diffusible chemical signals (\emph{i.e.},
\emph{chemotaxis}) and insoluble ECM molecules (\emph{i.e.},
\emph{haptotaxis}) described by $ E _{Field} = \mu \, c({\bf r})
L_xL_y, \;\; {\bf r}=(x,y)$ where $c(\bf r)$ is a local
concentration of particular species of signaling molecules in the
extracellular space and $\mu$ is an effective chemical potential.

Let $P({\bf r},{\bf L},t)$ denote the probability density for a
rectangular cell with its center of mass at ${\bf r}$  to have
dimensions ${\bf L}\equiv (L_x,L_y)$ at time $t.$ We use vectors
${\bf
e}_{1,2}$ to indicate changes in $x$ and $y$ dimensions: $
{\bf e}_1={\triangle r}(1,0), \ {\bf e}_2=\triangle r (0,1)$.
Let us normalize the total probability to the number of cells: $\int P({\bf
r},{\bf L},t)d{\bf r}d{\bf L}=N.$

Now assume that cells cannot occupy the same space.  This implies
that position ${\bf r}'$ and size ${\bf L}'$ of any neighboring cell
should satisfy the following excluded volume conditions: $ |x-x'|\ge
\frac{L_x+L_x'}{2},$ $ |y-y'|\ge \frac{L_y+L_y'}{2}.$

A discrete stochastic model of the cell dynamics under these
conditions is described by the following master equation
\begin{eqnarray} \label{pmasterxL1}
P({\bf r},{\bf L},t+\epsilon^2\triangle t) =\sum_{j=1}^2\Big \{\big
[\frac{1}{2}-\Phi_{j,l}({\bf r}-\frac{\epsilon}{2}{\bf
e}_j,{\bf L}+\epsilon{\bf e}_j; {\bf r},{\bf L}, t) \nonumber \\
-\Phi_{j,r}({\bf r}+\frac{\epsilon}{2}{\bf e}_j,{\bf L}+\epsilon
{\bf e}_j;{\bf r},{\bf L}, t)-T_l({\bf r}+\frac{\epsilon}{2}{\bf
e}_j,{\bf L}-\epsilon {\bf e}_j; {\bf r},{\bf L}, t)
\nonumber \\
-T_r({\bf r}-\frac{\epsilon}{2}{\bf e}_j,{\bf L}-\epsilon {\bf
e}_j;{\bf r},{\bf L}, t)\big ]P({\bf r},{\bf L},t)
\nonumber \\
+\Phi_{j,l}({\bf r},{\bf L};{\bf r}+\frac{\epsilon}{2}{\bf e}_j,
{\bf L}-\epsilon {\bf e}_j,t)P({\bf r}+\frac{\epsilon}{2}{\bf
e}_j,{\bf L}-\epsilon {\bf e}_j,t)
\nonumber \\
+\Phi_{j,r}({\bf r},{\bf L};{\bf r}-\frac{\epsilon}{2}{\bf e}_j,
{\bf L}-\epsilon {\bf e}_j,t)P({\bf r}-\frac{\epsilon}{2}{\bf
e}_j,{\bf L}-\epsilon {\bf e}_j,t)
\nonumber \\
+T_l({\bf r},{\bf L};{\bf r}-\frac{\epsilon}{2}{\bf e}_j,{\bf
L}+\epsilon {\bf e}_j,t)P({\bf r}-\frac{\epsilon}{2}{\bf e}_j, {\bf
L}+\epsilon {\bf e}_j,t)
\nonumber \\
+T_r({\bf r},{\bf L};{\bf r}+\frac{\epsilon}{2}{\bf e}_j,{\bf
L}+\epsilon {\bf e}_j,t)P({\bf r}+\frac{\epsilon}{2} {\bf e}_j, {\bf
L}+\epsilon {\bf e}_j,t) \Big \},
\end{eqnarray}
where $T_l({\bf r},{\bf L};{\bf r}',{\bf L}',t)$ and $T_r({\bf
r},{\bf L};{\bf r}',{\bf L}',t)$ denote probabilities of
transitions from a cell of length $L'$ and center of mass at $r'$
to a cell of dimensions $L$ and center of mass at $r$ without
taking into account excluded volume principle. (Terms with $T_l$
and $T_r$ in Eq. (\ref{pmasterxL1}) correspond to the case of
decreasing cell size $|{\bf L}|<|{\bf L}'|$ which justifies the
neglect of excluded volume.) $\Phi_{j,l}({\bf r},{\bf L};{\bf
r}',{\bf L}',t)$ and $\Phi_{j,r}({\bf r},{\bf L};{\bf r}',{\bf
L}',t)$ are probabilities of transitions taking into account
excluded volume.
 Subscripts $l$ and $r$ correspond to transitions by
addition/removal of a  row/colomn of pixels from the rear/lower
and front/upper ends of a cell respectively. According to the CPM
we have that $T_l(x,{\bf L};{\bf r}',{\bf L}')=T_r({\bf r},{\bf
L};{\bf r}',{\bf L}')= \frac{1}{8}\Phi\Big (E({\bf r},{\bf L})-
E({\bf r}',{\bf L}' ) \Big )$ where the factor of $1/8$ is due to
the fact that there are potentially 8 possibilities for increasing
or decreasing of $L_x$ and $L_y$.

We define  $\Phi_{j,l}({\bf r},{\bf L};{\bf r}',{\bf L}')\equiv
T_{l(r)}({\bf r},{\bf L};{\bf r}',{\bf L}')[1-\phi_{j,r(l)}({\bf
r},\, {\bf L},t)]$ where $\phi_{j,r(l)}({\bf r},\, {\bf L},t)$ is
the probability of another cell being in the immediate neighborhood
of a given cell and, therefore, preventing an increase of that
cells' length or width (excluded volume). We neglect triple and
higher order ``collisions" between cells resulting in the following
approximation formulas
\begin{eqnarray}\label{phidef1}
\phi_{1,k}({\bf r},\, {\bf L},t)=(N-1)(\epsilon\triangle r)^4
\nonumber \\
\times \sum\limits_{{\bf L}', y'} \Theta\big
(\frac{L_y+L'_{y}}{2}-|y'-y|\big ) P({\bf r}',{\bf L}',t) \Big
|_{x'=x+s\frac{L_x+L'_{x}}{2}}
\nonumber \\
\phi_{2,k}({\bf r},{\bf L},t)=(N-1)(\epsilon\triangle r)^4
\nonumber \\
\times \sum\limits_{{\bf L}',x'} \Theta\big
(\frac{L_x+L'_{x}}{2}-|x'-x|\big ) P({\bf r}',{\bf L}',t) \Big
|_{y'=y+s\frac{L_y+L'_{y}}{2}}
\end{eqnarray}
where $s= 1$ for $k=l$, $s=-1$ for $k=r$ and factor $N-1$ is due
to pairwise cell collisions.

We found by using Monte Carlo simulations (not shown) that solutions
of the master equation (Eq.$(\ref{pmasterxL1})$) with general
initial conditions quickly converge to $P({\bf r},{\bf
L},t)=P_{Boltz}({\bf r},{\bf L})p({\bf r},t)$ where $ P_{Boltz}({\bf
r},{\bf L})=Z({\bf r})^{-1}\exp(-\beta \triangle E_{length})$ is the
Boltzmann distribution and $\triangle E_{length}=E({\bf r},{\bf
L})-E_{min}=\lambda_x\tilde L_x^2+\lambda_y\tilde L_y^2+\tilde
L_x\tilde L_y\mu c({\bf r})$ and $\tilde {\bf L}={\bf L}-{\bf
L}^{(min)}$.
Also, $E_{min}=E({\bf r},{\bf L}^{(min)})$  is the minimal value
of the Hamiltonian (\ref{eq:2}) achieved at ${\bf L}={\bf
L}^{(min)}$ and
$ Z({\bf r})=(2\epsilon\triangle r)^2\sum\limits_{{\bf
L}}\exp(-\beta \triangle E_{length})\simeq
\frac{{2\pi}}{\beta\sqrt{4\lambda_x\lambda_y-\mu^2c({\bf r})^2}},
\ \epsilon\to 0 $ is an asymptotic formula for a  partition
function.

 The typical fluctuation of cell dimensions ($\tilde L_x,\
\tilde L_x$) are determined by $\beta\lambda_{x(y)} \tilde
L_{x(y)}^2\sim 1$. We now assume in addition that $ \beta
x_0^2\lambda_x \gg 1$ and $\quad \beta y_0^2\lambda_y \gg 1,$
where $x_0$ and $y_0$ are  typical scales of $P$ with respect to
$x$ and $y$. This means that $x_0\gg \tilde L_x, \ y_0\gg \tilde
L_y.$ We also assume that  the concentration of chemoattractant
$c({\bf r})$ is a slowly varying function of ${\bf r}$ on a scale
of the typical cell's length meaning that $ x_c/L_x\gg 1, \quad
y_c/L_y\gg 1,$ where $x_c$ and $y_c$ are typical scales for
variation of $c({\bf r})$ in $x$ and $y$.  We also make the
additional biologically relevant assumption that $
4\lambda_x\lambda_y\gg \mu^2c({\bf r})^2$
which means that change of typical cell size  due to chemotaxis
$(\delta L^{(chemo)}_x,\delta L^{(chemo)}_y)$ is small $|\delta
L^{(chemo)}_{x(y)}|\ll L^{(min)}_{x(y)}.$  Under all above
mentioned assumptions, the master Eq. $(\ref{pmasterxL1})$ is
transformed in the limit $\epsilon\ll 1$ into an
integro-differential equation describing evolution of the
probability density $p({\bf r},t)$ for the location of the
cellular center of mass
\begin{eqnarray}\label{pottscontinuous2dex}
\partial _tp=D_2\partial^2_{\bf r} p-\chi_0\partial_{\bf r} \cdot \big [p\, \partial_{\bf
r}c({\bf r})\big ]
\nonumber \\
  +\frac{D_2}{2}(N-1)\Big \{\partial_x [\psi_x
p]+\partial_y [\psi_y p] \Big \}
\nonumber \\
\psi_{x}=\int\limits^{y+L_y^{(min)}}_{y-L_y^{(min)}}\big [
p(x+L_x^{(min)},y')-p(x-L_x^{(min)},y')\big ] dy'
\nonumber \\
\psi_{y}=\int\limits^{x+L_x^{(min)}}_{x-L_x^{(min)}}\big [
p(x',y+L_y^{(min)})-p(x',y-L_y^{(min)}) \big ]dx'
\nonumber \\
\chi_0=-D_2\mu \beta L_{x}^{(min)}L_{y}^{(min)},
\end{eqnarray}
where $D_2=\frac{(\triangle r)^2}{16 \triangle t},
\partial^2_{\bf r}=\partial_x^2+\partial_y^2,$ $ \chi_0=-D_2\mu \beta L_{x}^{(min)}L_{y}^{(min)}, \,\,\,
L_{x}^{(min)}=L_{T_x}-\frac{J_{cm}}{\lambda_x}, \,\,\,
L_{y}^{(min)}=L_{T_y}-\frac{J_{cm}}{\lambda_y}$ and $\int p({\bf
r})d{\bf r}=N$. Lastly, we couple this equation to an equation
describing evolution of the external (chemotactic) field $c$
\begin{eqnarray}\label{ceq2D}
\partial_tc=D_c\partial^2_{\bf r}c-\gamma c +a\, p
\end{eqnarray}
where $D_{c}, \gamma$ and $a$ are diffusion, decay and production
rates of the field respectively. Note that the chemical is
produced by cells.

If excluded volume is not taken into account (i.e. assuming
$\psi_x=\psi_y=0$) Eqs. $(\ref{pottscontinuous2dex})$ and
$(\ref{ceq2D})$ reduce to the classical Keller-Segel system
\cite{KS} which has a finite time singularity and which was used
for modeling collapse (aggregation) of bacterial colonies
\cite{BrennerConstantinKadanoff1999}. Addition of excluded volume
significantly slows down collapse and, therefore, Eqs.
$(\ref{pottscontinuous2dex})$ and $(\ref{ceq2D})$ can be used for
simulating cellular aggregation for a much longer period of time.
Spongy bone formation considered in this paper, is accompanied by
secretion of a viscous or solid ECM (see below) which quickly
stabilizes a transient or metastable arrangement of cells into a
persistent microanatomy and therefore also prevents collapse.

\begin{figure}
  \begin{center}
    \begin{minipage}[c]{0.53\linewidth}
    \begin{center}
      \includegraphics[width=\linewidth]{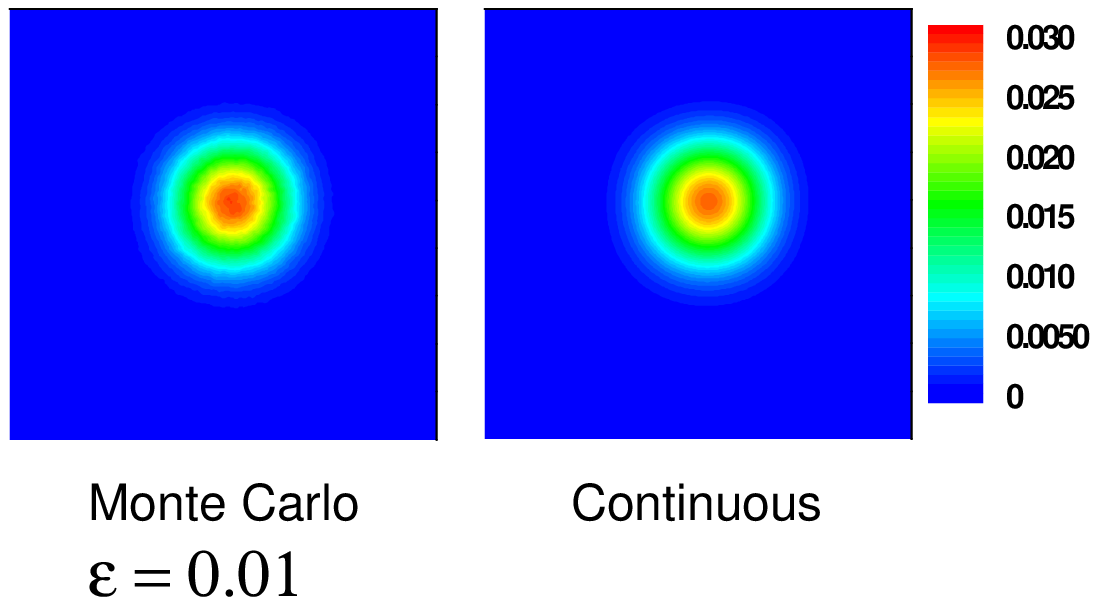}
      (a)
      \end{center}
    \end{minipage}\hfill
    \begin{minipage}[c]{0.45\linewidth}
    \begin{center}
          \includegraphics[width=\linewidth]{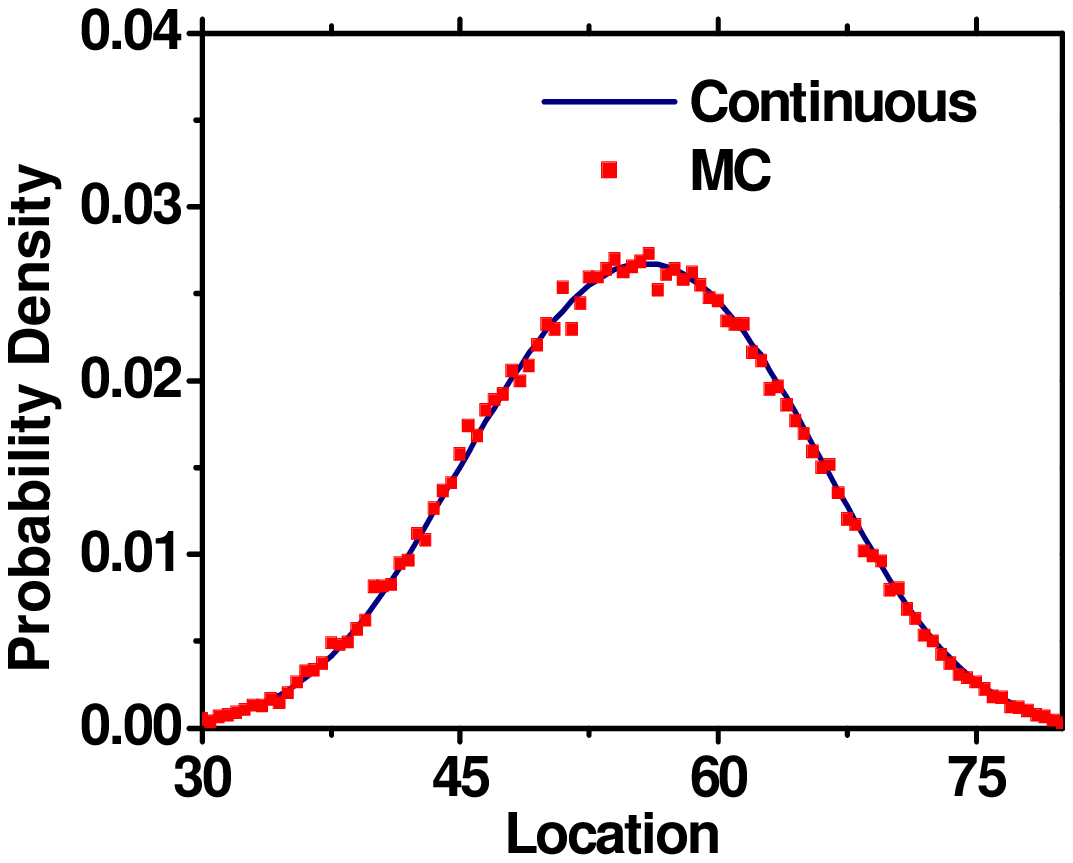}\\
          (b)
          \end{center}
  \end{minipage}
\caption{Comparison between mesoscopic CPM and macroscopic
continuous model. (a) Plot of a two-dimensional probability
density distributions for a CPM simulation of $12$ cells with
$\epsilon=0.01$ and numerical solution $p(x,y,t)$ of the
continuous Eq.$(\ref{pottscontinuous2dex})$. (b) Cross sections of
$p_{cpm}(x_0,y,t)$ and $p_{con}(x_0,y,t)$  at $x_0 = 53.0$ as
functions of $y$. \label{figMonConExdyn}}
  \end{center}
\end{figure}

Figure \ref{figMonConExdyn} demonstrates a very good agreement
between typical CPM simulation and numerical solution of the
continuous model $(\ref{pottscontinuous2dex})$ and
$(\ref{ceq2D})$. Both simulations were performed on a rectangular
domain $0\le x,y \le 100$ with simulation time $t_{end}=100$.
Parameters were chosen as follows: $\triangle r=1$,
$L_{T_x}=L_{T_y}=3$, $\lambda_x=\lambda_y=1.5$, $J_{cm}=2$,
$\beta=15$, $\mu=0.1$, $D_c=3.0$, $\gamma=0.00025$ and $a=0.2$.
 The time interval
between successive Monte Carlo steps was $\delta t=\epsilon^2
\triangle t=0.0001, \; \epsilon= 0.01$. Discrete form of the
equation ($\ref{ceq2D}$) was used to calculate the chemical field
dynamics on a $200\times200$ lattice with the time step $\Delta
t_c=0.0125$ and initial chemical field chosen in the form of
 $c_0(x,y)=\frac{(x-70)^2+(y-60)^2}{400}$. The typical size of
 the mesh used in the continuous
model was $1000\times1000$ and the time step was $0.002$.    A
large number of CPM simulations have been run to guarantee a
representative statistical ensemble. We assumed that at each time
step each cell released chemical content $a\Delta t_c$ which was
then distributed to four nearest chemical lattice sites.

\begin{figure}
\begin{center}
\includegraphics[width=0.8\linewidth]{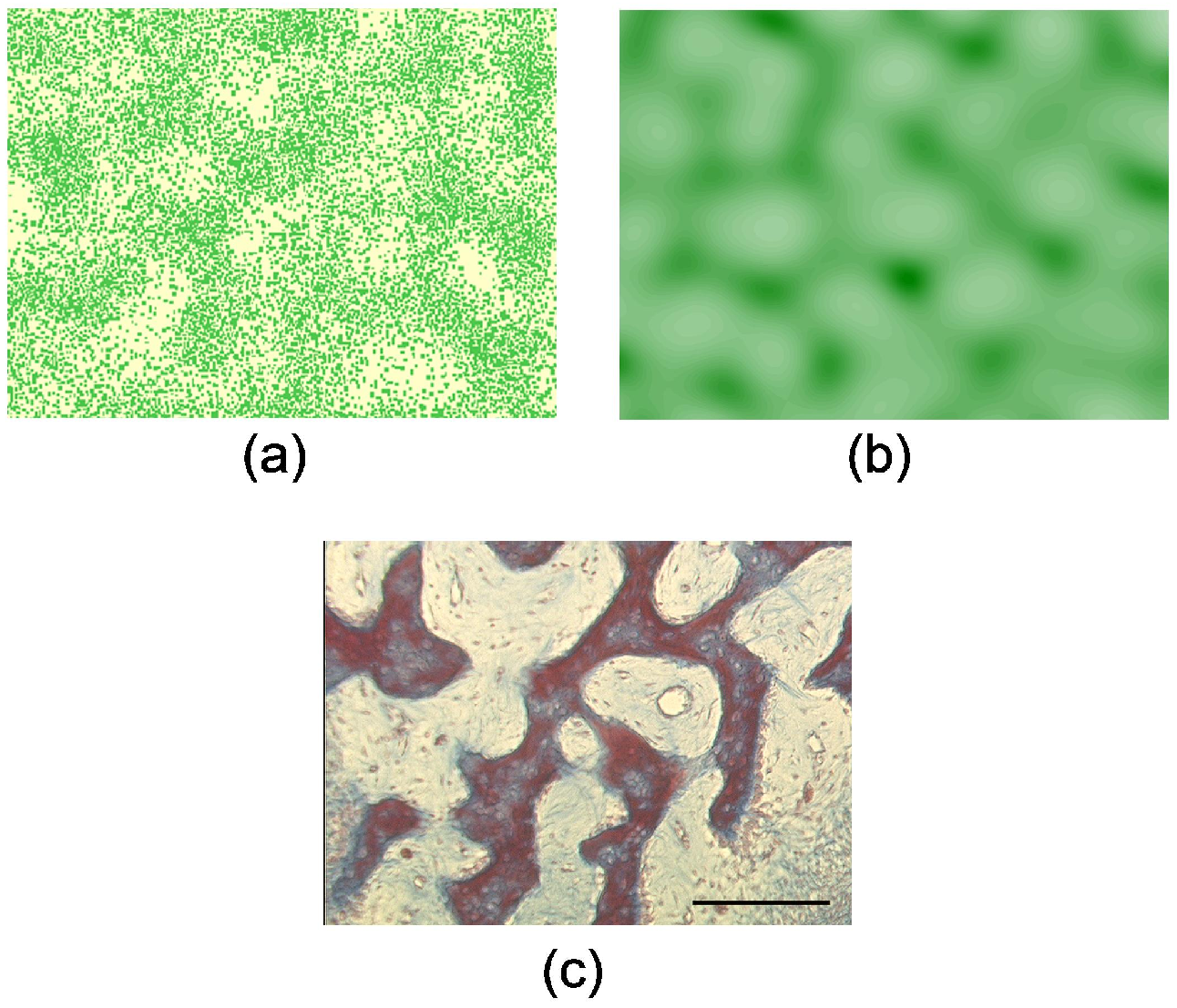}\\
\caption{Simulation of spongy bone formation process.  $\triangle
r=1$, $L_{T_x}=L_{T_y}=0.6$, $\lambda_x=\lambda_y=1.5$,
$J_{cm}=0.002$, $\beta=15$, $\mu=-0.1$, $D_c=0.5$, $\gamma=0.014$,
$\Delta t_c=\epsilon^2\Delta t=0.01, \epsilon= 0.1$, $t_{end}=180$.
(a) Monte Carlo CPM simulation. $a=0.7$. $ N=15000$ cells were
randomly distributed in a domain $0\le x,y \le 100$ with initial
chemical field at zero. (b) Numerical solution of the continuous
model resulting from a uniform initial cell density distribution and
with $5\%$ random fluctuation, $a=0.2$. (c) Histological section of
developing spongy bone in the rat skull. Trichrome stain.
Photographed from a section in the New York Medical College
Histology slide collection. The effective magnification of this
image is about 2x that of a and b. Scale bar: 100 micrometers.}
\label{figBone}
\end{center}
\end{figure}
In what follows, we illustrate the efficacy of the model by
applying it to the formation of spongy bone via the
intramembranous route. In this developmental phenomenon, which
generates portions of the skull, maxilla and mandible in
vertebrate organisms, bone cells, or osteoblasts, differentiate
directly from loosely packed mesenchymal cells.  The
differentiating cells secrete TGF-beta which acts chemotactically,
influencing cell migration while simultaneously inducing
production of ECM \cite{Kanaan}, which in developing bone is
termed osteoid \cite{cormack}.

Depending on local conditions, including initial cell density, the
bone will progress to a dense state or stop at a spongy state, in
which bony rods or trabeculae form a swiss-cheese-like network
(see Figure 3c) that eventually contains marrow tissue originating
from the circulation. Our mesoscopic and macroscopic model
simulations which start with initially dilute populations of cells
in a chemotactic field, subject to an excluded volume constraint,
result in a transiently appearing set of interconnected
multicellular trabeculae (see Figures 3a and 3b) similar to the
experimental picture (Figure 3c). In particular, in the
simulations and the developing tissue there are many nodes from
which three branches extend, but few with larger numbers.

In summary, we have derived a macroscopic continuous model
(\ref{pottscontinuous2dex}) from a mesoscopic two-dimensional CPM
with excluded volume constraint and coupled it to a model of
chemotaxis (\ref{ceq2D}). Numerical simulations confirm a very
good agreement between the CPM and macroscopic equations.
Numerical analysis of the macroscopic model facilitated
determination of conditions promoting formation of a lattice-like
aggregation pattern. This permitted us to locate the parameter
ranges within which the model cells in the CPM simulations behaved
qualitatively like the living cells that form multicellular
branches in spongy bone by intramembranous ossification(Figure
3c). In contrast to earlier suggestions that the trabecular
arrangement of spongy bone is based on pre-existing vascular
patterns \cite{Caplan}, or later-forming patterns of mineral
deposition \cite{courtin,tabor}, our results suggest that it can
arise from the self-organizing behavior of mesenchymal cells
interacting with their ECM.

This work was partially supported by NIH Grant No.
1R0-GM076692-01: Interagency Opportunities in Multiscale Modeling
in Biomedical, Biological and Behavioral Systems NSF 04.6071 and
NSF grants IBN-0344647, FIBR-0526854 and MRI DBI-0420980.

\end{document}